%% file: consum.tex
\documentclass[12pt,oneside]{article}

\usepackage{amsmath}
\usepackage{amsthm}
\usepackage{amssymb}
\usepackage{amsfonts}
\usepackage{array}
\usepackage{curves}
\usepackage{epic}
\usepackage{ifthen}

\newcommand{\N}{\mathbb{N}}

\newcommand{\R}{\mathbb{R}}

\newcommand{\Pb}{\mathbb{P}}
\newcommand{\E}{\mathbf{E}}

\newcommand{\1}{\mathbf{1}}

\newcommand{\F}{\mathcal{F}}

\newcommand{\Fb}{\mathbb{F}}

\newcommand{\ol}{\overline}

\newtheorem{theorem}{THEOREM}

\newtheorem{corollary}{COROLLARY}
\newtheorem{proposition}{PROPOSITION}
\newtheorem{lemma}{LEMMA}
\newtheorem{definition}{DEFINITION}

\newcounter{remark}
\newenvironment{remark}
{\refstepcounter{remark}
\vspace{4pt}
\noindent\textbf{Remark \arabic{remark}.}
}
{\vspace{4pt}}

\newcounter{example}\newsavebox{\comment}
\newenvironment{example}[1][]
{\refstepcounter{example}
\ifthenelse{\equal{#1}{}}{\sbox{\comment}{\text{}}}{\sbox{\comment}{\textbf{ (#1)}}}
\vspace{4pt}\noindent\textbf{Example \arabic{example}\usebox{\comment}.}
}
{\vspace{4pt}}

\def\bi{\begin{itemize}}
\def\ei{\end{itemize}}

\def \yr#1{\overline{#1}\vrule\ }
\def\Axn#1#2{A_{#1:\yr{#2} } }

\begin{document}


\title{Consumption processes and positively homogeneous projection properties}
\author{Tom Fischer\thanks{The author gratefully acknowledges funding by a research grant from The 
Actuarial Profession in the UK. Thanks are owed to an anonymous referee whose suggestions helped 
to considerably improve the paper. -- Author's address:
Maxwell Institute for Mathematical Sciences, and Actuarial Mathematics and Statistics, 
Heriot-Watt University, Edinburgh EH14 4AS, United Kingdom. Tel: +44 131 451 4371,
Email: {\tt fischer@ma.hw.ac.uk}.
}\\
{Heriot-Watt University, Edinburgh}} 
\date{\today}
\maketitle

\begin{abstract}
We constructively prove the existence of time-discrete consumption processes for stochastic 
money accounts that fulfill a pre-specified positively homogeneous projection 
property (PHPP) and let the account always be positive and exactly zero at the 
end. One possible example is consumption rates forming a martingale under the 
above restrictions.  For finite spaces, it is shown that any strictly 
positive consumption strategy with restrictions as above possesses at least one 
corresponding PHPP and could be constructed from it.  We also consider numeric 
examples under time-discrete and -continuous account processes, cases with infinite 
time horizons and applications to income drawdown and bonus theory.
\end{abstract}






\input{body}


\end{document}

%% file: body.tex
\section{Introduction}

The purpose of this paper is twofold. On the one hand, it introduces a
general method to describe and construct consumption processes
for stochastic money accounts where the consumption processes at the same time
fulfill certain restrictions, e.g.~non-negativity of the money account and total
consumption in the end. This method is general in the sense that at
least in the case of finite spaces it can be shown that basically any meaningful
consumption process can be described and constructed by it, even if the corresponding 
process is originally obtained by a completely different method, e.g.~by expected 
utility maximization. On the other hand, the considered method itself allows 
to directly describe stochastic properties of consumption processes. These 
processes can then be constructed fulfilling at the same time the above 
restrictions. An example of such a property could be that the consumption 
rates form a martingale. This could be understood as a ``smooth" form of consumption.
In contrast to the economic approach of deriving consumption processes from maximization 
of expected utility, it will be explained below in what sense the direct postulation of stochastic 
properties for the consumption rates could be seen as an actuarial approach. 
Applications to actuarial problems like bonus theory and stochastic annuities 
will show the usefulness of the proposed methods. For instance, a ``smooth", 
i.e.~martingale bonus strategy for closed with-profits funds will be derived.
Another example explains a benchmark procedure for determining upper consumption
limits for pension policies under so-called income drawdown.

To be more specific, let us consider the following situation.
A private or institutional agent invests an initial amount of currency units 
in a money account. We assume that without consumption the value of a positive 
amount of money in the account always stays strictly positive and develops
in a non-deterministic way.
The investor wants to take money regularly, e.g.~yearly, out of the account.  
There are certain restrictions for the consumption procedure.
Typically, the money account will not be allowed to become negative at any time, money 
will only be consumed and not added to the account,
and after a certain period of time, e.g.~10 years, all funds will have to be
consumed. Except for the consumption, the investor is not able to influence
the dynamics of the account. For instance, if the money account represented
a stock market index, we would assume that the investor is ``small", and the consumption
rates would therefore not influence the prices, i.e.~the dynamics of the account
between two successive consumption dates.
There also is no alternative investment opportunity for our agent. Otherwise, he or she
could influence the overall investment dynamics. This, however, we do not wish since
such influence is not possible in the case of many real consumption situations.
Under the assumptions made so far, one might be interested in the following questions:

(1) How could the investor specify desirable properties of the (in general
non-deterministic) consumption process?

(2) Having described such properties and knowing the dynamics of the money account without
consumption, does there exist a consumption process with these properties?

(3) If the second question was answered with ``yes", how could the
corresponding consumption process be constructed and pursued?

Regarding the first question, there exists a vast variety of ways in 
which such properties could be described. For instance, and first of all, we could think of the
already mentioned
utility-optimizing approaches. Utility-based methods play for good reasons a predominant role in 
economic theory.
One could mention applications in portfolio optimization (selection by expected utility optimization), 
utility-hedging and insurance premium calculation (utility indifference pricing). However, alternatives
exist, and they are widely used. To mention some, we could think of portfolio performance or risk 
optimization with respect to measures like RORAC, Value-at-Risk or Expected Shortfall. For hedging
we could think of quadratic and quantile hedging approaches. Talking of insurance premiums, 
calculation principles like the Expectation and the Variance Principle are commonly used. 
All mentioned alternatives have in common that a rather direct stochastic 
property or quantity - like (minimizing) a quantile, a variance, an expectation - is used 
to obtain a result like optimal risk or performance, a premium etc. 
Since premium calculation is supposedly the oldest financial application of such stochastic 
properties, (note: a risk measure can always be interpreted as a premium principle), it might 
be justified to call these alternatives ``actuarial".

For the consumption problem, less work has been carried out on non-utility alternatives. 
In some way this is
astonishing since especially in actuarial contexts consumers and/or companies seem to have great 
interest in ``smoothness", e.g.~in the sense of ``smooth" consumption or bonus rates. However, bonus 
rates for instance are often declared in
an {\em ad hoc} manner, and it is often not at all clearly defined what ``smooth" is supposed to 
mean. Usually, it is understood to mean that rates from one year to the next are changing not too 
abruptly, i.e.~in some way ``smoothly".
For the following, let us denote the absolute consumption rate at time 
$k$ by $X_k$ $(k = 0, 1,2,\ldots, K)$. 
As an example of a ``smooth" consumption process, one might consider rates that form a 
martingale, i.e.~$X_k = \E[X_{k+1}|\F_k]$ where $\F_k$ denotes the information
up to time $k$. In other words,
``consume every year what you can expect to consume in the following year".
Since in the deterministic limit case a martingale is a constant function,
this is perhaps as ``smooth" as a discrete time process with certain restrictions in place
can get. It is for these reasons that the martingale property is one of the prime examples in 
this paper.

The martingale property is local in the sense that a rate at time $k$, $X_k$, is obtained 
from a projection, or operator, applied to its successor at $k+1$, $X_{k+1}$. In the course
of the paper it will turn out that this feature is extremely useful for the definition, 
for the proof of the existence and for the construction of consumption
processes under the earlier mentioned restrictions. In particular,
we will consider consumption processes that possess what we call a 
{\em positively homogeneous projection property} (PHPP). 
A PHPP means that for any $k<K$ the rate $X_k$ is determined by its successor 
$X_{k+1}$ in the way that $X_k = \psi_k(X_{k+1})$, with $\psi_k$ being an operator
taking care of the right measurability of $X_k$ with respect to the available information at $k$.
We assume that $\psi_k$ maps non-negative (non-positive) random variables to again
non-negative (non-positive) ones. Furthermore, $\psi_k$ is assumed to be positively homogeneous 
in the sense that 
$\psi_k(YX) = Y\psi_k(X)$ for $X$ $\F_{k+1}$-measurable and $Y\geq 0$ and $\F_k$-measurable.
The final rate $X_K$ is determined by a possibly random fraction $0\leq \psi_K \leq 1$ of 
what is left in the account at 
time $K$. For instance, $\psi_K \equiv 1$ would mean that all funds are consumed at the end.
Examples of $\psi_k$ that are considered later are multiples of 
conditional expectations (``consume every year $x$\% of what you can expect to consume 
in the following year",
e.g.~$x=100$ would imply a martingale) and conditional quantiles (``consume this year $X_k$, 
and with probability 75\% you are allowed to consume more next year").
A submartingale strategy (e.g.~choosing $x=95$ in the conditional expectation case) might be 
considered as to provide ``smooth growth". As outlined before, institutional investors,
in particular insurance companies, might be interested in such consumption solutions,
especially in the ``smooth" martingale consumption strategy.
For insurance companies, the earlier mentioned applications to bonus
theory and income drawdown should be of special interest.
An additional benefit of the above examples is that they are comparatively easy 
to explain to customers or policyholders.
While the communication of features of an insurance policy that come from
a utility approach to a policyholder with a non-mathematical background might be difficult, 
most people seem to have a natural imagination of what an expectation (average) or 
a probability (percentage) is.

The main result of the paper shows that for any PHPP, $\psi = (\psi_0,\ldots,\psi_K)$, and any 
strictly positive stochastic process, $S$, for the money account without consumption, there exists a 
consumption process, $X = (X_k)_{k = 0, \ldots, K}$, which possesses this property, never adds money to 
the account (i.e.~is non-negative) and never lets the money account with consumption, $A$, become 
negative. We call an $X$ with the latter two properties {\em regular}. We give a constructive proof 
which enables us to explicitly calculate strategies in many cases, e.g.~when we have a tree model 
for $S$, or for certain PHPPs when $S$ has i.i.d.~growth factors
(e.g.~log-L\'evy processes like geometric Brownian motions).
Although the projection properties seem to be quite exclusive at first, it can be shown for
finite probability spaces that {\em any} strictly positive regular consumption process 
(i.e.~restrictions as above and no zero-consumption) possesses at least one such property 
and could therefore be constructed from it. In other words, the proposed method is not restricted
to a certain type of consumption processes. It rather represents a general mathematical concept
for the construction of arbitrary consumption processes fulfilling the usual restrictions.

To compare the results of this paper with the existing literature seems to be somewhat difficult.
Most papers regarding (optimal) consumption (see e.g.~Merton (1971), Karatzas, Lehoczky 
and Shreve (1987), and Cox and Huang (1989)) work with utility frameworks, and the 
questions asked and/or solved are different from the ones addressed above. 
For example, the consumption problem is usually defined as the objective to maximize
expected utility having at the same time the opportunity to invest in different kinds of 
assets. As we neither consider a utility framework, nor have an objective for optimization,
our considerations are very different from this kind of investigation. Furthermore,
we do not consider a portfolio problem as only one investment opportunity is given
for the reasons indicated above.  

The proof of the main result described above shows that for a given value process $S$ a PHPP 
$\psi$ uniquely defines a regular consumption process $X(S,\psi)$ which can be determined by a 
combined backward-forward recursion. Although this method is useful, for instance, when dealing 
with a tree model for $S$, its applications are limited since
the number of paths might grow exponentially in time.
Furthermore, the process $S$ could be an almost arbitrary
time-continuous process making such a calculation generally impossible.
For these reasons, the possibility to simulate $X(S,\psi)$ would be helpful. 
However, assuming one can simulate paths of $S$, it is obvious that this is possible when 
relative consumption rates are deterministic (by a relative rate of consumption we mean the 
absolute consumption rate expressed as a fraction of the actual value of the money 
account at that time).
The second part of the paper therefore takes a closer look at such strategies,
derives more explicit results, and shows the implications for underlying PHPPs of the 
corresponding absolute consumption process. Examples illustrate that deterministic relative 
consumption rates
are of foremost interest when $S$ has i.i.d.~growth factors (e.g.~log-L\'evy processes).
In this part of the paper we also consider cases of consumption for infinite time horizons.
Finally, we explain in two actuarial examples how the results of the paper can be applied to 
pensions with so-called income drawdown, and to what extent they can be applied to bonus 
strategies for closed with-profits funds.

The paper is organized as follows.
In Section \ref{2}, we start with basic considerations of absolute consumption rates.
The relationship between the consumption process, the money account process without consumption
and the money account with consumption is derived. Section \ref{okidoki} explores relative rates
of consumption. In Section \ref{4}, PHPPs are introduced, and the central theorem of the paper
proves the existence of consumption strategies with such properties. 
Section \ref{detrel} derives more explicit results for cases in which the relative
consumption rates are deterministic. 
The section contains (partially numeric) examples of explicit consumption solutions
when a binomial tree and a geometric Brownian motion model the dynamics of the money account 
without consumption.
In Section \ref{detpet}, we consider consumption for infinite time horizons
when relative consumption rates are again deterministic. 
Section \ref{actuarial} is on the mentioned example of income drawdown. Section \ref{bonus}
is on bonus for closed with-profits funds. In Section \ref{conclude}, we conclude.


\section{Absolute rates of consumption}
\label{2}

Given the positive real numbers $\R^+_0$ as time axis, let $S=(S_t)_{t\geq 0}$ be a strictly 
positive stochastic process adapted to a filtration $\Fb=(\F_t)_{t\geq 0}$ which is defined
on an underlying probability space $(\Omega, \F_\infty, \Pb)$. We consider $S$ 
as the value process of a stochastic {\em money account without consumption}. For instance,
$S$ could represent the development of an initial investment of $S_0 = 100,000$ currency units 
into stocks, an index or a fund at time 0. $S$ describes the dynamics of the value of this 
investment under the assumption that no money is taken out or paid in after time 0. As usual,
the $\sigma$-algebra $\F_t$ represents all available information up to time $t$.

At integer times $k\in\N_0=\{0,1,2,\ldots\}$, a consumer intends to take money out of the account.
The amount taken out at time $k$ is denoted by $X_k$, the {\em absolute rate of consumption}
at time $k$. The real-valued
process $X=(X_k)_{k\in\N_0}$, which we call the {\em consumption process},
is adapted with respect to the filtration $(\F_k)_{k\in\N_0}$, so to speak
a sub-filtration of $(\F_t)_{t\geq 0}$, which provides the information up to integer
times, only. 
We now consider the value process of the money account {\em with} consumption.
Our basic assumption here is that after the consumption of rate $X_k$, the 
process of the account is a rescaled version of the value process one would have without 
having consumed this particular rate. 
This is straightforward. In practice one would, for example, consume 10 
shares out of a stock of 100 identical shares of a certain fund.
Furthermore, we assume that the consumer is a ``small investor'', i.e.~the volume
of all consumption taking place is small in relation to the total trading volume
of the market. We can therefore assume that consumption does not influence market prices.

Let $A = (A_t)_{t\geq 0}$ denote the value process of the money account with consumption.
We have (in a mathematical sense, define) 
\begin{equation}
\label{alder}
A_t = 
\begin{cases}
S_0 &  \text{if } t=0,\\
(A_k-X_k)S_t/S_k &  \text{if } k<t\leq k+1\; (k\in\N_0).
\end{cases}
\end{equation} 
Figure 1 illustrates this relationship between $S$, $X$, and $A$.
It is clear that $(A_t)_{t\geq 0}$ is an adapted process w.r.t.~$(\F_t)_{t\geq 0}$.
The discrete time process $(A_k)_{k\in\N_0}$ represents the value in the money account 
immediately {\em before} consumption at times $k\in\N_0$. From a mathematical point of
view, it is not necessary to make any continuity assumptions for $S$ in the following.
However, if $S$ is left-continuous, so is $A$. Provided left-continuity of $S$, 
\eqref{alder} should be interpreted such that consumption takes place immediately 
{\em before} a possible jump in $S$ at the same time. In the case of right-continuity of 
$S$, the interpretation should be that consumption takes place immediately 
{\em after} a possible jump in $S$ at the same time.

\begin{figure}
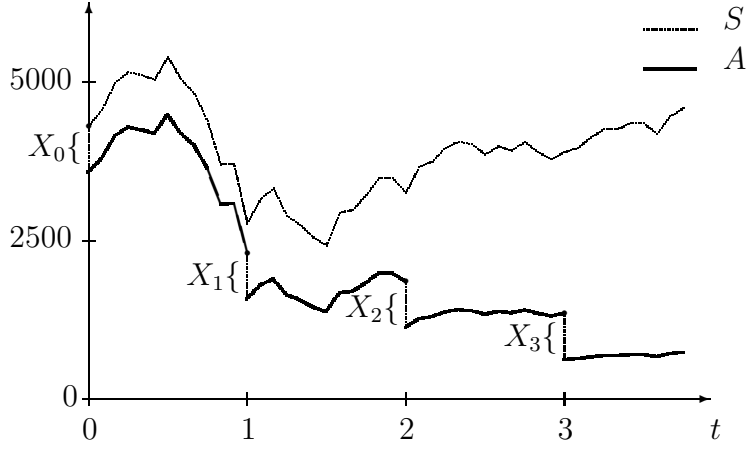

\begin{center}
\vspace{5mm} 
\include{picture1}
\end{center}
\caption{Sample paths for $(S_t)_{t\geq 0}$ and $(A_t)_{t\geq 0}$}
\end{figure}

\begin{lemma}
\label{waah}
For $k\in\N_0$,
\begin{equation}
\label{alderr}
A_t = \left(1-\sum_{i=0}^{k}\frac{X_i}{S_i}\right)S_t, \quad k<t\leq k+1.
\end{equation}
\end{lemma}

\begin{proof}
By induction. \eqref{alderr} is true for $k=0$.
Now, assume \eqref{alderr} for some 
$k \in \N_0$. For $k+1<t\leq k+2$,
\begin{eqnarray}
\label{alderen}
A_t & \stackrel{\eqref{alder}}{=} & (A_{k+1}-X_{k+1})S_t/S_{k+1}\\
& \stackrel{\eqref{alderr}}{=} &
\left(\left(1-\sum_{i=0}^{k}\frac{X_i}{S_i}\right)S_{k+1} - X_{k+1}\right)S_t/S_{k+1}\\
\nonumber & = & \left(1-\sum_{i=0}^{k+1}\frac{X_i}{S_i}\right)S_t .
\end{eqnarray}
  \end{proof}

It is clear from \eqref{alder} that
\begin{equation}
\label{g8}
0 \leq X_k \leq (<)\, A_k , \quad k\in\N_0,
\end{equation} 
implies
\begin{equation}
\label{g8+2}
A_t \geq (>)\, 0 , \quad t>0.
\end{equation} 
Note that \eqref{g8} also implies that once all the money has been consumed,
i.e.~once one has had $X_{k_0} = A_{k_0}$ for some
$k_0 \in\N_0$, the values of $X_{k}$ for $k > k_0$ and the values of
$A_t$ for $t>k_0$ are zero.

\begin{definition}
{\em 
A consumption process $X=(X_k)_{k\in\N_0}$ fulfilling \eqref{g8} is called {\em regular}
with respect to $S$. We call $X$ {\em strictly regular} w.r.t.~$S$ if
\begin{equation}
\label{g8+3}
0 <\, X_k <\, A_k , \quad k\in\N_0.
\end{equation}
}
\end{definition}

The following result on scaling will be used in Section \ref{bonus}.
It follows directly from \eqref{alder}.

\begin{lemma}
\label{rescaling}
Let $(F_t)_{t\geq 0}$ be a strictly positive adapted process, and let $X$ be a 
consumption process w.r.t.~$S$ ($S$ as before). Define for $t\geq 0$
\begin{equation}
\tilde{S}_t = F_tS_t \quad\text{and}\quad \tilde{X}_t = F_tX_t ,
\end{equation}
and let $\tilde{A}_t$ be defined w.r.t.~$\tilde{S}$ and $\tilde{X}$ 
in analogy to \eqref{alder}. Under these assumptions,
\begin{equation}
\label{skye}
\tilde{A}_t = F_tA_t, \quad t\geq 0,
\end{equation} 
and $\tilde{X}$ is regular w.r.t.~$\tilde{S}$ if and only if $X$ is regular w.r.t.~$S$.
 
\end{lemma}


\section{Relative rates of consumption}
\label{okidoki}

So far we have described consumption by means of absolute consumption rates. This is reasonable
as in the following we will be interested in properties of the consumption process $X$. However, 
some consumption properties under consideration can be described more easily 
when considering relative consumption rates. By a {\em relative rate of consumption} we mean 
the absolute consumption rate expressed as the percentage of the value of the money account at the 
time of consumption. For a regular consumption process $X$ w.r.t.~$S$, we therefore call
\begin{equation}
\label{easy0}
Z_k = 
\begin{cases}
r\in\R & \text{ if } A_k=0\\
\frac{X_k}{A_k} & \text{ else}
\end{cases}
\end{equation}
a relative consumption rate belonging to the absolute consumption rate $X_k$. 
We easily obtain the following lemma, whose proof is left to the reader.
\begin{lemma}
\label{wee}
Any adapted process $Z = (Z_k)_{k\in\N_0}$ with 
\begin{equation}
0\leq (<)\, Z_k \leq (<)\, 1 , \quad k\in\N_0,
\end{equation}
determines by $X_k=Z_kA_k$ a (strictly) regular consumption process w.r.t.~$S$.
For $k<t\leq k+1$ $(k\in\N_0)$,
\begin{equation}
\label{cold}
A_t = \left(\prod_{i=0}^{k}(1-Z_i)\right)S_t, 
\end{equation}
and
\begin{equation}
\label{easy2}
X_k = Z_k\left(\prod_{i=0}^{k-1}(1-Z_i)\right)S_k .
\end{equation}
 
\end{lemma}
Note that for regular consumption, \eqref{alderr} and \eqref{cold} imply
\begin{equation}
\label{easy4}
\prod_{i=0}^{k}(1-Z_i) = \left(1-\sum_{i=0}^{k}\frac{X_i}{S_i}\right), \quad k\in\N_0.
\end{equation}


\section{PHPPs for finite time horizons}
\label{4}

Consider a finite time axis $[0,K]\subset \R$ where $K\in\N_0$. One could consider processes
$S= (S_t)_{0\leq t \leq K}$, $X=(X_k)_{k = 0, \ldots, K}$, and $A= (A_t)_{0\leq t \leq K}$
similar to their counterparts on the infinite time axis. It is easy to check that all
statements made and derived above are also true when restricted to such a finite 
time horizon. For instance, a consumption process regular w.r.t.~$S= (S_t)_{0\leq t \leq K}$
is then a process $X=(X_k)_{k = 0, \ldots, K}$ 
adapted to the discrete time filtration $(\F_k)_{k = 0, \ldots, K}$ fulfilling 
\eqref{g8} for $k = 0, \ldots, K$. 

We now describe operators that will help us to define what we will call 
a {\em positively homogeneous projection property} (PHPP) of a consumption process $X$.

Let $L^0(\F_k)$ $(k = 0, \ldots, K)$ denote the set of real-valued $\F_k$-measurable random variables,
and let $L_{+}^0(\F_k)$ ($L_{-}^0(\F_k)$) denote the cone of non-negative (non-positive) 
$\F_k$-measurable random variables. For $k=0,\ldots,K-1$, 
let $\psi_k$ be a mapping
\begin{equation}
\label{psi_1}
\psi_k: L^0(\F_{k+1}) \longrightarrow L^0(\F_k) ,
\end{equation}
such that
\begin{eqnarray}
\label{psi_3}  L_+^0(\F_{k+1}) & \stackrel{\psi_k}{\longrightarrow} & L_+^0(\F_k),\\
 \label{psi_33} L_-^0(\F_{k+1}) & \stackrel{\psi_k}{\longrightarrow} & L_-^0(\F_k),
\end{eqnarray}
and
\begin{equation}
\label{psi_2}
\psi_k(YX) = Y\psi_k(X)
\end{equation}
for $X\in L^0(\F_{k+1})$ and $Y\in L_+^0(\F_k)$. Furthermore, let
\begin{equation}
\label{psi_4}
\psi_K \in L_+^0(\F_K) \text{ such that } \psi_K \leq 1.
\end{equation}
Note that $\psi_K$, in contrast to $\psi_0,\ldots,\psi_{K-1}$, is defined as a random variable 
and not as an operator.
Property \eqref{psi_2} implies for any $F\in\F_k$ that $\psi_k(X)|_F$ only depends on $X|_F$
since we can choose $Y = \1_F$ in \eqref{psi_2}. 
The operators $\psi_k$ in \eqref{psi_1} are not necessarily projections in
the usual geometric sense. Neither do they have to be linear, nor is necessarily
$\psi_k\circ\psi_k = \psi_k$. For $|\F_K|=\infty$ it might not always be possible
to define $\psi_0,\ldots,\psi_{K-1}$ on the whole $L^0(\F_1),\ldots,L^0(\F_{K})$.
If this is the case, one can often restrict the considered spaces and ranges the operators
and properties are defined on, such that most of the results showed in the following are still 
applicable.

\begin{theorem}
\label{theo_1}
An adapted process $(X_k)_{k = 0, \ldots, K}$ is non-negative and fulfills the conditions
\begin{equation}
\label{prop_1} 
X_k = \psi_k(X_{k+1}) ,\quad k = 0, \ldots, K-1,
\end{equation}
and
\begin{equation}
\label{prop_2} 
X_K = \psi_K A_K
\end{equation}
if and only if
\begin{equation}
\label{XZ}
X_k = Z_kA_k ,\quad k = 0, \ldots, K,
\end{equation}
where
\begin{equation}
\label{Z_2}
Z_{k-1} = \frac{\psi_{k-1}(Z_kS_k)}{S_{k-1}+\psi_{k-1}(Z_kS_k)} , \quad k = 1, \ldots, K,
\end{equation}
and
\begin{equation}
\label{Z_1}
Z_K = \psi_K .
\end{equation}
Furthermore, the uniquely determined process $X$ is a regular consumption process w.r.t.~$S$.
\end{theorem}

Note that \eqref{Z_2} is a pointwise equation, i.e.~we actually define $Z_{k-1}(\omega)$
for $\omega \in \Omega$, but leave out the argument for convenience.

\begin{proof}
First some observations on regularity and non-negativity of $X$.  
A non-negative consumption process $X$ fulfilling \eqref{prop_1} and \eqref{prop_2} must be regular.
If $X$ was not regular, there would be a smallest $k$ such that 
$X_k > A_k \geq 0$ on some $F\in \F_k$. For $k=K$, \eqref{prop_2} would be a contradiction
since $\psi_K \leq 1$.
For $k<K$, this would imply $A_l < 0$ on $F$ for $l = k+1, \ldots, K$ since $X$
is non-negative. From \eqref{psi_33} and \eqref{psi_2} we would obtain 
\begin{equation}
\label{last}
X_k\1_F = \psi_k(\psi_{k+1}(\ldots \psi_{K-1}(\psi_KA_K\1_F)\ldots )) \leq 0, 
\end{equation}
contradicting $X_k>0$ on $F$. 
For non-negative, regular $X$, \eqref{prop_1} and \eqref{prop_2} imply $A> 0$, i.e.~$A_t>0$ for
$0\leq t \leq K$. To prove this observe that since $S>0$, $A>0$ is equivalent
to $(A_k)_{k = 1, \ldots, K} >0$. If $(A_k)_{k = 1, \ldots, K}$ was zero anywhere, there would 
exist some $k\in\{0,\ldots,K-1\}$
and $F\in \F_k$ such that $X_k = A_k > 0$ on $F$. Since $A_K$ had to be zero on $F$ as well,
\eqref{last} would lead to the contradiction $X_k\1_F = 0$.
Regarding the second part of the theorem, 
observe that $Z=(Z_k)_{k = 0, \ldots, K}$ as given by \eqref{Z_2} and \eqref{Z_1}
is a non-negative process adapted to $(\F_k)_{k = 0, \ldots, K}$.
From the definition of the $\psi_k$ and from \eqref{Z_2} and \eqref{Z_1}, it is clear that
$0\leq Z_K \leq 1$ and
\begin{equation}
\label{regul}
0\leq Z_k < 1 , \quad k = 0, \ldots, K-1.
\end{equation}
Therefore, any consumption process $X$ with \eqref{XZ} is regular.\\
``If": Assume \eqref{XZ}, \eqref{Z_2} and \eqref{Z_1}. \eqref{prop_2} follows
from \eqref{XZ} and \eqref{Z_1}. From \eqref{Z_2} and \eqref{regul} we obtain 
$\psi_{k-1}(Z_kS_k) = \frac{S_{k-1}Z_{k-1}}{1-Z_{k-1}}$ for $k = 1, \ldots, K$.
By Lemma \ref{wee}, for $k = 1, \ldots, K$,
\begin{align}
\label{nasowas}
\psi_{k-1}(X_k) &
\stackrel{\eqref{XZ}}{=}
\psi_{k-1}(Z_kA_k) \stackrel{\eqref{cold}\&\eqref{psi_2}}{=} \psi_{k-1}(Z_kS_k)\prod_{i=0}^{k-1}(1-Z_i)\\
\nonumber& \;=  \frac{S_{k-1}Z_{k-1}}{1-Z_{k-1}}\prod_{i=0}^{k-1}(1-Z_i)   
= S_{k-1}Z_{k-1}\prod_{i=0}^{k-2}(1-Z_i) \\
\nonumber & \stackrel{\eqref{cold}}{=} Z_{k-1}A_{k-1} \stackrel{\eqref{XZ}}{=} X_{k-1}.
\end{align}
``Only if" is obtained by backward induction.
For $k=K$, \eqref{XZ} with \eqref{Z_1} follows from \eqref{prop_2}. From $k$ to
$k-1$ we get as follows.
\begin{eqnarray}
X_{k-1} & = & \psi_{k-1}(X_k) = \psi_{k-1}(Z_kA_k)\\
\nonumber & \stackrel{\eqref{alder}}{=} & \psi_{k-1}(Z_kS_k)\frac{A_{k-1}-X_{k-1}}{S_{k-1}}.
\end{eqnarray}
Therefore,
\begin{align}
X_{k-1} & = \frac{\psi_{k-1}(Z_kS_k)}{S_{k-1}+\psi_{k-1}(Z_kS_k)} A_{k-1}.
\end{align}
  \end{proof}

\begin{definition}
\label{scalinv}
Define $\psi = (\psi_0,\ldots,\psi_K)$ where $\psi_0,\ldots,\psi_K$ are as above.
We say that a consumption process $X$ for a value process $S$
has the {\em positively homogeneous projection property (PHPP)} $\psi$ if it fulfills
\eqref{prop_1} and \eqref{prop_2}.
We denote such an $X$ by $X(S, \psi)$, and the corresponding $Z$ by $Z(S, \psi)$.
\end{definition}

\begin{corollary}
\label{Weinheim}
For any non-negative $X(S,\psi)$ one has for $k=0,\ldots,K-1$ that $Z_k<1$, or equivalently $X_k<A_{k}$.
Furthermore, $A_t >0$ for $0\leq t \leq K$. 
\end{corollary}

\begin{proof}
See proof of Theorem \ref{theo_1}.
  \end{proof}

Theorem \ref{theo_1} shows that given some value process $S$ and 
a PHPP $\psi$, there exists one and only one non-negative consumption 
process $X(S,\psi)$ with this property. Furthermore, this $X(S,\psi)$ is regular.
The theorem provides us with a backward-forward recursion scheme for $X$.
First, by backward recursion, the process $Z(S,\psi)$ of the relative consumption rates
can be determined (Eq.~\eqref{Z_2} and \eqref{Z_1}). Then, $Z$ is used in
the forward recursion formula \eqref{easy2}, or, by \eqref{easy4}, in
\begin{equation}
\label{easy6}
X_k = Z_k\left(1-\sum_{i=0}^{k-1}\frac{X_i}{S_i}\right)S_k, \quad k = 0, \ldots, K.
\end{equation}

The number of examples of positively homogeneous projection properties is vast.
For instance, if $(\psi_0,\ldots,\psi_K)$ is a PHPP, any $(B_0\psi_0,\ldots,B_K\psi_K)$
for $\F_k$-measurable $B_k$ ($0\leq B_k\leq 1$, $k = 0, \ldots, K$) is one, too. 
In the following, we briefly consider two particular classes.
More detailed examples (a binomial tree model and a geometric Brownian motion 
for $S$) are considered in Section \ref{detrel}. Note that the following two
examples could be numerically implemented for any reasonably sized finite 
stochastic tree modelling $S$.

\begin{example}[Conditional expectations]
\label{ex1}
Assume $|\F_K|<\infty$. For $k=0,\ldots,K-1$, $X\in L^0(\F_{k+1})$ and $c>0$ let
\begin{equation}
\label{condexp}
\psi_k(X) = c\E[X|\F_k]
\end{equation} 
and $\psi_K \equiv d$, $0\leq d\leq 1$. It is easy to check that \eqref{psi_1} to
\eqref{psi_4} are fulfilled. The consumption process with the corresponding PHPP
fulfills $X_k = c\E[X_{k+1}|\F_k]$ for $k=0,\ldots,K-1$ and $X_K = dA_K$.
A consumer following this strategy consumes at each time $k$ exactly $c\times 100\%$ of
what he or she can
expect to consume one step ahead at time $k+1$. The final rate $X_K$ would consume 
$d\times 100\%$ of the capital left in the account at time $K$. Note that $c=1$ means that
$X$ is a martingale, $c<1$ implies a submartingale, i.e.~ the consumer 
can expect growing consumption rates.
\end{example}

\begin{example}[Conditional quantiles]
\label{ex2}
Assume $|\F_K|<\infty$ and set $\psi_K \equiv d$, $0\leq d\leq 1$, and
\begin{equation}
\label{condquant}
\psi_k(X) = q_\alpha(X|\F_k) ,
\end{equation} 
where $q_\alpha(X|\F_k)$ denotes a conditional quantile.
For instance, for $X_k = q_{0.4}(X_{k+1}|\F_k)$, the consumption rate at $k+1$ 
would be higher than the one at $k$ with probability $60\%$. 
The standard definition of an $\alpha$-quantile ($\alpha\in [0,1]$) being
\begin{equation}
\label{quantile}
q_\alpha(X) = \inf\{y: \Pb(X\leq y) \geq \alpha\},
\end{equation} 
one might choose to define
\begin{equation}
\label{condquant2}
q_\alpha(X|\F_k) = \text{ess inf}\{Y: Y \in L^0(\F_k)\text{ and } \E[\1_{X\leq Y}|\F_k] 
\geq \alpha\text{ a.s.}\},
\end{equation} 
where $\E[\1_{X\leq Y}|\F_k]$ is the probability of $\{X\leq Y\}$ conditional on $\F_k$. 
In the finite case, \eqref{condquant2} is equivalent to
\begin{equation}
\label{condquant3}
q_\alpha(X|\F_k) = \sum_{F\in \mathcal{A}_k} \1_F\inf\{y: \Pb(X\leq y|F) \geq \alpha\},
\end{equation} 
where $\mathcal{A}_k$ denotes the partition of $\Omega$ consisting of the atoms of $\F_k$.
Note that \eqref{condquant} is an example of a non-linear (non-additive) PHPP.
\end{example}

The examples above were restricted to $|\F_K| < \infty$ since 
a definition of the considered operators on the whole $L^0(\F_K)$ is generally
not possible. However, consideration of consumption in more general spaces 
using these properties is often still possible when restricting the considered 
spaces in a reasonable way, e.g.~working on integrable subspaces in the 
conditional expectations case, and when the dynamics of $S$ are suitable.
The elements that a restricted support of the 
operators $\psi_{k-1}$ $(k=1,\ldots,K)$ should definitely contain in order to make 
Theorem \ref{theo_1} work are $Z_kA_k$ and $Z_kS_k$ ($Z_k$ as in Theo.~\ref{theo_1}). 
The minimal functionality the operators should have (besides
mapping non-negative/-positive functions to again non-negative/-positive ones)
would be given by $\psi_{k-1}(Z_kA_k) = \psi_{k-1}(Z_kS_k)\prod_{i=0}^{k-1}(1-Z_i)$. 
If these minimal requirements are satisfied, the proof of Theo.~\ref{theo_1} shows that 
$X(S,\psi)$ exists.\\

It is remarkable that not only every positively homogeneous projection property leads to a 
regular consumption process, but also the inverse is true under certain restrictions.
 
\begin{proposition}
\label{aubacke} 
Assume $|\F_K|<\infty$. Let $X$ be a regular consumption process such that $(X_k)_{k=0,\ldots,K-1}$
strictly regular and $X_K>0$.
Then there exists a PHPP $\psi$ such that
$X = X(S,\psi)$. This statement is in general not true for only regular
consumption processes.
\end{proposition}

\begin{proof}
Let
\begin{equation}
\psi_K = X_K/A_K,
\end{equation} 
and for $k = 0, \ldots, K-1$ and $Y\in L^0(\F_{k+1})$ 
\begin{equation}
\label{eigudewie}
\psi_k(Y) = \frac{X_{k}\E[Y|\F_{k}]}{\E[X_{k+1}|\F_{k}]} .
\end{equation} 
Clearly, these $\psi_k$ fulfill \eqref{psi_1} to \eqref{prop_1}. 
To show that we can not find a positively homogeneous projection property $\psi$
for every regular consumption process, it is sufficient to take a strictly regular one,
called $X$ here, e.g.~obtained from the relative consumption rates $Z_0 = \ldots = Z_K = 0.5$,
select a time $k_0$ ($0<k_0\leq K$) and define $Y_k = X_k$
for $k<k_0$ and $Y_k = 0$ for $k\geq k_0$. $Y= (Y_k)_{k=0,\ldots,K}$ is now a regular consumption
process, but has no corresponding PHPP since this would lead to the contradiction 
$0 < Y_{k_0-1}=\psi_{k_0-1}(Y_{k_0}) = \psi_{k_0-1}(0) =0$.
  \end{proof}

The relation between the strictly positive regular consumption processes (as in Proposition
\ref{aubacke}) and PHPPs is no bijection.
For instance, \eqref{eigudewie} is linear, but we have seen that there are genuinely non-linear 
examples of PHPPs, e.g.~conditional quantiles. 
The fact that we can find a PHPP for any strictly positive regular consumption process might 
be useful for the construction of such a process fulfilling a certain additional property (e.g. being
optimal w.r.t.~some objective like an expected utility).
If from this latter property the corresponding positively homogeneous projection property could 
be derived, we could construct the process by backward-forward recursion as outlined before.


\section{Deterministic relative consumption rates}
\label{detrel}

The results of the previous section have shown that for a value process $S$ a given
positively homogeneous projection property uniquely defines a regular consumption process.
This process can be determined by a combined backward-forward recursion.
It has been mentioned in the introduction that this method is useful when dealing with 
tree models for $S$, but the applications are limited since
the number of paths might grow too fast.
Also, $S$ could be an (almost) arbitrary time-continuous process such 
that calculations like these are generally impossible. Therefore,
the possibility to simulate the consumption process $X(S,\psi)$ would be 
very useful.

From the results of the previous section, we see that if $Z$ itself is 
deterministic, i.e.~$Z_k\equiv z_k \in [0,1]$, $k=0,\ldots,K$, formula \eqref{easy2} provides 
a direct way to calculate $X$ -- provided one can simulate paths of $S$. 
In this section we will have a closer look at such strategies and will show
that in many cases deterministic relative consumption strategies can be interpreted as 
absolute consumption derived from a particular positively homogeneous projection property.
An example illustrates that these concepts are of foremost interest when $S$ 
is a log-L\'evy process, for instance a geometric Brownian motion.

\begin{proposition}
\label{relative2}
Let $a_0, \ldots, a_K$ be a family of non-negative real numbers. For any PHPP $\psi$ with 
\begin{equation}
\label{agga1}
\psi_{k}(S_{k+1}/S_{k}) \equiv a_{k} , \quad k=0,\ldots,K-1,
\end{equation}
and
\begin{equation}
\label{agga2}
\psi_{K} \equiv a_K
\end{equation}
one has for $Z(S,\psi)$
\begin{equation}
\label{XX1}
Z_k \equiv \frac{\prod_{j=k}^{K}a_{j}}{1+\sum_{j=k}^{K-1}\prod_{i=j}^{K}a_{i}}, 
\quad k = 0, \ldots, K.
\end{equation}
Furthermore, for $X(S,\psi)$ one has that for $k = 0, \ldots, K$ 
\begin{eqnarray}
\label{agga4} X_k & = & \frac{\prod_{j=k}^{K}a_{j}}{1+\sum_{j=k}^{K-1}\prod_{i=j}^{K}a_{i}}A_k\\
\label{agga3} & = & \frac{\prod_{j=k}^{K}a_{j}}{1+\sum_{j=0}^{K-1}\prod_{i=j}^{K}a_{i}}S_k . 
\end{eqnarray}
\end{proposition}

\begin{proof}
We prove \eqref{XX1} by backward induction. Assume \eqref{agga1} and \eqref{agga2}.
For $k=K$, \eqref{XX1} is true since $Z_K = \psi_K = a_K$.
Now suppose \eqref{XX1} is true for some $k>0$. Then
\begin{eqnarray}
\label{guggug}
\psi_{k-1}(Z_{k}S_k) & \stackrel{\eqref{psi_2}}{=} & 
\left(\frac{\prod_{j=k}^{K}a_{j}}{1+\sum_{j=k}^{K-1}\prod_{i=j}^{K}a_{i}}\right)
S_{k-1} \cdot\psi_{k-1}\left(\frac{S_{k}}{S_{k-1}}\right)\\
\nonumber & = & 
\left(\frac{\prod_{j=k-1}^{K}a_{j}}{1+\sum_{j=k}^{K-1}\prod_{i=j}^{K}a_{i}}\right)
S_{k-1} .
\end{eqnarray}
Therefore,
\begin{equation}
\label{guggugg}
Z_{k-1} = \frac{
\left(\frac{\prod_{j=k-1}^{K}a_{j}}{1+\sum_{j=k}^{K-1}\prod_{i=j}^{K}a_{i}}\right)S_{k-1}}{S_{k-1}+ 
\left(\frac{\prod_{j=k-1}^{K}a_{j}}{1+\sum_{j=k}^{K-1}\prod_{i=j}^{K}a_{i}}\right)S_{k-1}}
= \left(\frac{\prod_{j=k-1}^{K}a_{j}}{1+\sum_{j=k-1}^{K-1}\prod_{i=j}^{K}a_{i}}\right) .
\end{equation}
Observe that \eqref{agga4} follows directly from \eqref{XX1} and \eqref{XZ}.
To prove \eqref{agga3} we again utilize induction. Obviously, \eqref{agga3} holds for $k=0$.
Now, assuming \eqref{agga3} to be valid up to $k$, we obtain from \eqref{agga4}
and \eqref{alderr}

\begin{eqnarray}
X_{k+1} & = & \left(1-\sum_{i=0}^{k}\frac{X_i}{S_i}\right) 
\left(\frac{\prod_{j=k+1}^{K}a_{j}}{1+\sum_{j=k+1}^{K-1}\prod_{i=j}^{K}a_{i}}\right)
S_{k+1} \\
\nonumber & = & \left(1-\sum_{i=0}^{k}
\left(\frac{\prod_{j=i}^{K}a_{j}}{1+\sum_{j=0}^{K-1}\prod_{h=j}^{K}a_{h}}\right) 
\right) 
\left(\frac{\prod_{j=k+1}^{K}a_{j}}{1+\sum_{j=k+1}^{K-1}\prod_{i=j}^{K}a_{i}}\right)
S_{k+1} \\
\nonumber & = & \left(\frac{1+\sum_{j=k+1}^{K-1}\prod_{i=j}^{K}a_{i}}{1+\sum_{j=0}^{K-1}\prod_{i=j}^{K}a_{i}}\right)
\left(\frac{\prod_{j=k+1}^{K}a_{j}}{1+\sum_{j=k+1}^{K-1}\prod_{i=j}^{K}a_{i}}\right)
S_{k+1} \\
\nonumber & = & 
\left(\frac{\prod_{j=k+1}^{K}a_{j}}{1+\sum_{j=0}^{K-1}\prod_{i=j}^{K}a_{i}}\right) 
S_{k+1}  .
\end{eqnarray}
  \end{proof}

The next result inverts Proposition \ref{relative2} in some way.

\begin{proposition}
\label{relative1}
Let a regular consumption process $X(S,\psi)$ with PHPP $\psi$ have corresponding 
deterministic relative consumption rates $Z(S,\psi)$ with
\begin{equation}
\label{zz}
Z_k \equiv z_k, \quad  (k = 0, \ldots, K)
\end{equation}
where $z_k\in [0,1)$, $k = 0, \ldots, K-1$, and $z_K\in [0,1]$. Let $k_0$ be the
maximal $k$ such that $z_k = 0$. Then,
\begin{equation}
\label{portobello}
\psi_k(S_{k+1}/S_k) \equiv \frac{z_k}{(1-z_k)z_{k+1}}, \quad k = k_0, \ldots, K-1,
\end{equation}
and
\begin{equation}
\label{portobello2}
\psi_K \equiv z_K.
\end{equation}
\end{proposition}

\begin{proof}
$Z$ is given by \eqref{Z_2} and \eqref{Z_1}. 
Note that \eqref{Z_2} implies $Z_0 = \ldots = Z_{k_0} \equiv z_{k_0} = 0$.
We first prove \eqref{portobello} and
\eqref{portobello2} for $k_0$ equal to $K$ or $K-1$.
If $k_0=K$, then $\psi_K\equiv z_K = 0$. 
If $k_0=K-1$, then $\psi_K\equiv z_K$ and $z_{K-1}=0$.
From \eqref{Z_2} and \eqref{psi_2} we derive $\psi_{K-1}(S_K)=0$, and by \eqref{psi_2}
$\psi_{K-1}(S_{K}/S_{K-1}) = \psi_{K-1}(S_{K})/S_{K-1} = 0$.
For $k_0<K-1$, \eqref{Z_2} implies inductively that
$\psi_{k-1}(S_k/S_{k-1})$, $k=k_0+1,\ldots,K$, is constant and strictly positive. 
From \eqref{Z_2} we easily derive \eqref{portobello}.
  \end{proof}

\begin{corollary}
\label{relative5}
Assume $|\F_K|<\infty$.\\
$(i)$ There exists for any 
family $a_0,\ldots,a_K$ of real numbers with $a_k\geq 0$ $(k=0,\ldots,K-1)$, and $a_K\in [0,1]$ a 
positively homogeneous projection property $\psi$ such that \eqref{agga1} and \eqref{agga2}
hold.\\
\label{relative3}
$(ii)$ For any family $z_0,\ldots,z_K$ of real numbers fulfilling  
$z_k\in [0,1)$ for $k = 0, \ldots, K-1$, $z_K\in [0,1]$, and 
$z_k=0$ for $k = 0, \ldots, k_0$ (where $k_0$ is as in Prop.~\ref{relative1}), there exists
a positively homogeneous projection property $\psi$ such that for $Z(S,\psi)$ 
one has $Z_k \equiv z_k$, $k = 0, \ldots, K$.
\end{corollary}

Corollary \ref{relative5} and Proposition \ref{relative2} and \ref{relative1} answer the question in
what way deterministic relative consumption can be interpreted as 
absolute consumption under specific positively homogeneous projection properties. For strictly
positive regular consumption processes, the mere existence of at least one corresponding PHPP
follows already from Prop.~\ref{aubacke}. The following proof of the corollary makes use of
Prop.~\ref{relative2}.

\begin{proof}
$(i)$ Define $\psi_K = a_K$ and for $k = 0, \ldots, K-1$
\begin{equation}
\label{yeah}
\psi_k(X) = \frac{a_{k}S_{k}\E[X|\F_{k}]}{\E[S_{k+1}|\F_{k}]}
\end{equation} 
These $\psi_k$ fulfill \eqref{psi_1} to \eqref{psi_4}, as well as \eqref{agga1} and 
\eqref{agga2}.\\
$(ii)$ Define a family of positive real numbers $a_0,\ldots,a_K$ with
\begin{equation}
\label{portobello3}
a_k = \frac{z_k}{(1-z_k)z_{k+1}}, \quad k = k_0, \ldots, K-1,
\end{equation}
and $a_K=z_K$. Note that $a_0,\ldots,a_{k_0-1}$ can be arbitrarily chosen.
Apply now part $(i)$ and Prop.~\ref{relative2}. From \eqref{XX1} one obtains for 
$k = k_0, \ldots, K-1$
\begin{eqnarray}
Z_k & \equiv & \frac{\prod_{j=k}^{K}a_{j}}{1+\sum_{j=k}^{K-1}\prod_{i=j}^{K}a_{i}} 
= \frac{z_K\prod_{j=k}^{K-1} \frac{z_j}{(1-z_j)z_{j+1}} }{1+\sum_{j=k}^{K-1}
z_K\prod_{i=j}^{K-1} \frac{z_i}{(1-z_i)z_{i+1}} }\\
\nonumber & = & \frac{\frac{z_k}{\prod_{j=k}^{K-1} (1-z_j)} }{1+\sum_{j=k}^{K-1}
\frac{z_j}{\prod_{i=j}^{K-1} (1-z_i)}  }\\
\nonumber & = & z_k \left(\prod_{j=k}^{K-1} (1-z_j) + \sum_{j=k}^{K-1}
z_j\prod_{i=k}^{j-1} (1-z_i) \right)^{-1} = z_k .
\end{eqnarray}
Since $a_{k_0}=0$, we obtain $Z_k\equiv 0$ ($k = 0, \ldots, k_0$) from \eqref{XX1}.
  \end{proof}

\begin{corollary}
\label{specialcol}
Assume in Prop.~\ref{relative2} that $a_0=\ldots=a_{K-1}=a$. Then, for $k = 0, \ldots, K$,
\begin{equation}
\label{XX4}
X_k = \left(\frac{a_Ka^{K-k}}{1+a_K\sum_{j=1}^{K-k}a^j}\right) A_{k} 
= \left(\frac{a_Ka^{K-k}}{1+a_K\sum_{j=1}^{K}a^j}\right) S_{k} .
\end{equation}
 
\end{corollary}

\begin{remark}
\eqref{agga3} and \eqref{XX4} enable us to directly calculate the consumption 
process $X(S,\psi)$, provided $S$ can be simulated. Note that \eqref{agga3} and \eqref{XX4}
also imply that the relative standard deviation of $X_k$  and $S_k$ are the same, i.e.
$\mathbf{SD}[X_k]\big/ \E[X_k] = \mathbf{SD}[S_k]\big/ \E[S_k]$ ($\mathbf{SD}$ denoting the
standard deviation). In other words, the fluctuations of $X_k$  and $S_k$ are identical in
relative terms.
\end{remark}

If the discrete time process $(\log S_k)_{k=0,\ldots,K}$ derived from $S$ has 
i.i.d.~increments, then all $\frac{S_k}{S_{k-1}}$, $k=1,\ldots,K$,
are i.i.d.~and strictly positive. Geometric Brownian motions, or generally log-L\'evy 
processes would be classical examples of such $S$. From previous results,
especially Prop.~\ref{relative2} and the PHPPs in the Examples \ref{ex1} and \ref{ex2}, 
it seems obvious that such processes for $S$ are suitable candidates for applications.

\begin{example}[Binomial tree]
\label{yeahh}
Let us work with a binomial tree for $(S_k)_{k=0,\ldots,K}$, 
i.e.~all $S_k/S_{k-1}$ are i.i.d.~($k = 1, \ldots, K$),
and $\Pb[S_k/S_{k-1}=u_1]=p$ and $\Pb[S_k/S_{k-1}=u_2]=1-p$, where $0<u_1<u_2$ and 
$p\in (0,1)$.
Consider now Example \ref{ex1}. Clearly, we are here in a situation where Corollary
\ref{specialcol} can be applied with 
\begin{equation}
\psi_k(S_{k+1}/S_k) \equiv c(pu_1+(1-p)u_2) = a
\end{equation}
and $\psi_K=a_K=d$. Eq.~\eqref{XX4} calculates paths of $X$ for known paths of $(S_k)_{k=0,\ldots,K}$.
Calculations can now be carried out by standard spreadsheet programs. Note that we do not
need to separately calculate each path of the tree. The outcomes at times $k=0,\ldots,K$ are
sufficient, i.e.~calculations do not grow exponentially, but in second power, only.
Table \ref{tab:1} shows two paths of $(S_k)_{k=0,\ldots,K}$ ($S_0 = 10,000$) and the corresponding
consumption process $X$ for $K=10$, $c=1/1.02$, $d=1$, $u_1 = 1.02$, $u_2 = 1.06$, and $p=0.5$.
This particularly means that we have a submartingale with total consumption in the end. 
The value of $c$, that means an expected annual growth of $X$ by 2\%, is chosen such that even 
the worst case scenario, $(u_1,u_1,\ldots,u_1)$,
provides a slightly increasing rate of consumption. 
\end{example}

\begin{table}
\caption{Example \ref{yeahh} (binomial tree)}
\label{tab:1}       
\begin{tabular}{l|rrr|rrr}
\hline\noalign{\smallskip}
& \multicolumn{3}{l}{worst path} & \multicolumn{3}{l}{best path}\\
$k$ & $S_k$ & $X_k$ & $A_k$ & $S_k$ & $X_k$ & $A_k$\\
\noalign{\smallskip}\hline\noalign{\smallskip}
0 & 10,000.00 & 999.90 & 10,000.00 & 10,000.00 & 999.90 & 10,000.00\\
1 & 10,200.00 & 1,000.28 & 9,180.11 & 10,600.00 & 1,039.51 & 9,540.11\\
2 & 10,404.00 & 1,000.66 & 8,343.42 & 11,236.00 & 1,080.69 & 9,010.64\\
\ldots & \ldots & \ldots & \ldots & \ldots & \ldots & \ldots\\
8 & 11,716.59 & 1,002.98 & 2,951.44 & 15,938.48 & 1,364.38 & 4,014.94\\
9 & 11,950.93 & 1,003.36 & 1,987.43 & 16,894.79 & 1,418.43 & 2,809.59\\
10 & 12,189.94 & 1,003.75 & 1,003.75 & 17,908.48 & 1,474.63 & 1,474.63\\
\noalign{\smallskip}\hline
\end{tabular}
\end{table}

\begin{figure}
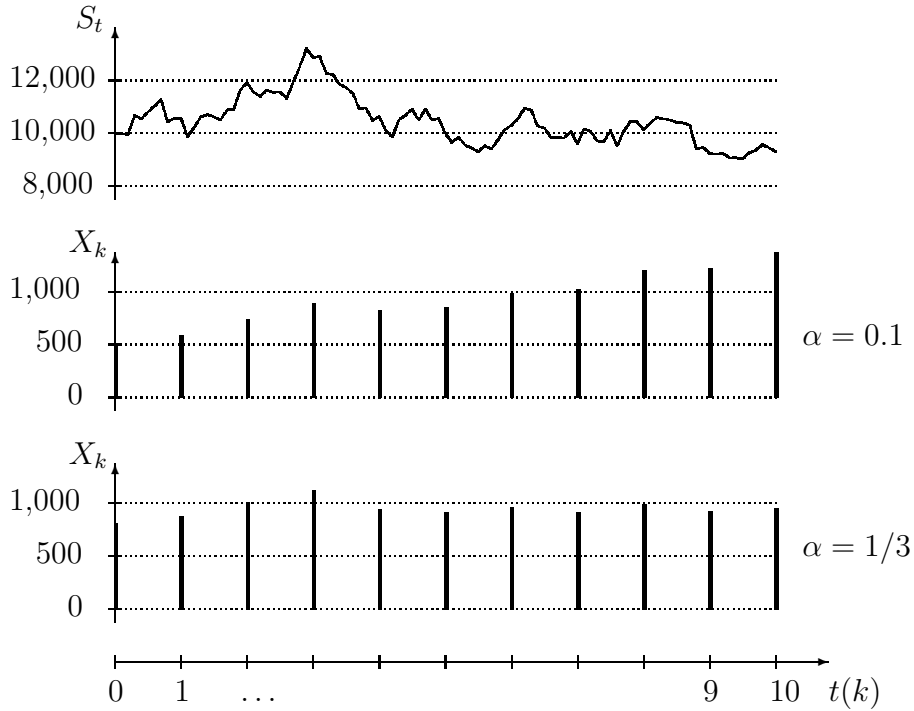

\begin{center}
\include{picture2}
\end{center}
\caption{Example \ref{GBM} (geometric Brownian motion). Sample paths for $S$ and $X$ ($K=10$)
under the conditional quantile property.}\label{fig:2}
\end{figure}

\begin{example}[Geometric Brownian motion]
\label{GBM}
We reconsider Example \ref{ex2} with $S$ being a geometric Brownian motion,
i.e.~
\begin{equation}
S_t = S_0\exp(\mu t+\sigma W_t), \quad S_0,\mu,\sigma >0,
\end{equation} 
where $W=(W_t)_{t\geq 0}$ denotes a 
standard Brownian motion. Let $\F=(\F_t)_{t\geq 0}$ be the filtration generated by $W$.
We have 
\begin{equation}
S_k/S_{k-1} = \exp(\mu + \sigma Y_k),
\end{equation}
where $Y_k = W_{k}-W_{k-1} \sim N(0,1)$ independent from $\F_{k-1}$ ($k = 1, \ldots, K$).
From the definition of $q_\alpha$ in Example \ref{ex2}, it is not difficult to derive that
$q_\alpha(S_k/S_{k-1}|\F_{k-1}) \equiv q_\alpha(S_k/S_{k-1})$. 
Assume now that $\psi_K \equiv 1$ and for $k = 0, \ldots, K-1$ 
\begin{equation}
\label{Mia}
\psi_k(X) = q_\alpha(X|\F_k),
\end{equation}
where \eqref{Mia} is only considered for $X$ where $q_\alpha(X|\F_k)$ is measurable.
Although the $\psi_k$ are now not defined on the whole $L^0(\F_{k+1})$, it is easy
to check that Corollary \ref{specialcol} still applies. 
For instance, we could choose $d=1$ (total consumption at the end), $S_0=10,000$, 
$\mu=0.02$, $\sigma = 0.1$ and $\alpha = 0.1$ (alternatively, $\alpha = 1/3$). We obtain
$a = q_{0.1}(S_k/S_{k-1}) = \exp(0.02 + 0.1 \Phi^{-1}(0.1)) \approx 0.897488$ 
$(a\approx 0.977191)$.
Eq.~\eqref{XX4} provides us with values of $X$ when $S$ is simulated
by standard Monte-Carlo methods. Figure \ref{fig:2} shows for $K=10$ a sample path for the 
corresponding money account without consumption, $S$. The corresponding consumption 
rates for $\alpha=0.1$ ($\alpha =1/3$) are plotted.
Pursuing the outlined strategy, the probability of being allowed to
consume more one year later is always 90\% (66.67\%). The chosen path shows an unpleasant
scenario in which the money account is less worth after ten years than in the beginning.
The graphs for $X$ illustrate how these kind of consumption strategies can cope with 
such situations.
\end{example}

\begin{remark}
Using Corollary \ref{specialcol}, it is easy to derive that for a fixed interest scenario with
$S_t=S_0\times (1+i)^t$, $i>-1$, $S_0\equiv const >0$, the martingale property 
(Ex.~\ref{ex1}) with $c=d=1$ implies that 
\begin{equation}
X_k = \frac{S_0}{\ddot a_{\ol{K+1}|}} , \quad k = 0, \ldots, K.
\end{equation}
Here, $\ddot a_{\ol{K+1}|}$ denotes in international actuarial notation the present value of an 
annuity-certain of 1 currency unit annually in advance for $K+1$ years. E.g., for $i=4\%$ and $S_0 = 10,000$, 
one has $\ddot a_{\ol{K+1}|} = 9.1109$ and therefore $X_0=\ldots =X_K = 1097.59$. 
This illustrates why $X_0$ calculated for a martingale strategy for a stochastic account could 
be interpreted as a stochastic generalisation of the benefit of an annuity-certain with present
value $S_0$.
One could also interpret $S_0/X_0$ as the generalisation of $\ddot a_{\ol{K+1}|}$, the capital
at time 0 that is necessary to have an (expected) annual benefit of 1.
\end{remark}


\section{Cases of perpetual consumption}
\label{detpet}

In this section, we will examine deterministic relative consumption rates
for infinite time horizons. As an application, we could think of 
the Nobel Prize Committee that has to make sure
that sufficiently stable amounts of money can be drawn from an existing fund every 
year, with no time limit being known. 

For an infinite time horizon, we denote a
consumption process with respect to $S$ by $X(S,\psi)$ if $X_k = \psi_k(X_{k+1})$, $k\in\N_0$,
where $\psi_k$, $k\in\N_0$, satisfies \eqref{psi_1} to \eqref{psi_2}.
We then say that $X$ has the positively homogeneous projection property $\psi=(\psi_0,\psi_1,\ldots)$.

\begin{proposition}
\label{theo_3}
Let $\{a_k: k\in\N_0\}$ be a family of strictly positive real numbers and assume that $\psi$ is
a positively homogeneous projection property with
\begin{equation}
\label{boy1}
\psi_{k}(S_{k+1}/S_{k}) \equiv a_{k}, \quad k\in\N_0.
\end{equation}
There exists a strictly regular consumption process $X(S,\psi)$ fulfilling 
\begin{equation}
\label{boy2}
X_k = z_kA_k , \quad k\in\N_0,
\end{equation}
for a family $\{z_k: k\in\N_0\}$ of real numbers if and only if
\begin{equation}
\label{boy2.1}
0 < z_0 ,
\end{equation}
\begin{equation}
\label{boy3}
z_k = z_0\left/\left(\prod_{i=0}^{k-1}a_i - z_0\sum_{i=0}^{k-1}\prod_{j=i}^{k-1}a_j\right)\right. ,
\quad k\in\N_0,
\end{equation}
and
\begin{equation}
\label{boy4}
z_0 < \frac{\prod_{i=0}^{k-1}a_i}{1 + \sum_{i=0}^{k-1}\prod_{j=i}^{k-1}a_j}, \quad k\in\N_0.
\end{equation}
\end{proposition}

Observe that \eqref{boy4} implies $z_0< 1$.

\begin{proof}
``Only if'': We obtain
\begin{align}
\label{boy7}
X_k \stackrel{\eqref{boy2}}{=} z_kA_k & = \psi_k(X_{k+1}) \stackrel{\eqref{boy2}}{=} 
\psi_k(z_{k+1}A_{k+1})\\
\nonumber  & \stackrel{\eqref{alder}}{=} z_{k+1}\psi_k([A_k-z_kA_K]S_{k+1}/S_k)\\ 
\nonumber  & = z_{k+1}\psi_k([1-z_k]A_kS_{k+1}/S_k)\\ 
\Rightarrow \quad\frac{z_k}{(1-z_k)z_{k+1}} & = \psi_{k}(S_{k+1}/S_{k}) ,
\end{align}
and hence
\begin{equation}
\label{boy5}
z_{k+1} \stackrel{\eqref{boy1}}{=} \frac{z_k}{(1-z_k)a_{k}}, \quad k\in\N_0.
\end{equation}
In \eqref{boy5}, no division by zero can occur since strict regularity of $X$ implies 
$0 < z_k < 1$ , $k\in\N_0$. We now show \eqref{boy3} and \eqref{boy4} for $k\in\N_0$ by induction. 
Assume for any $k\in\N_0$ \eqref{boy3} and \eqref{boy4}.
Observe that \eqref{boy3} is valid for $k=0$, and 
\eqref{boy4} implies that the denominator in \eqref{boy3} is greater than
$z_0$, which is greater than $0$. Now,
\begin{eqnarray}
\label{boy6}
z_{k+1} = \frac{z_k}{(1-z_k)a_{k}} & = &  
\frac{z_0}{\left(\prod_{i=0}^{k-1}a_i - z_0\sum_{i=0}^{k-1}\prod_{j=i}^{k-1}a_j 
- z_0 \right)a_k}\\
\label{maeh^3} & = & \frac{z_0}{\prod_{i=0}^{k}a_i - z_0\sum_{i=0}^{k}\prod_{j=i}^{k}a_j} .
\end{eqnarray}
From $z_{k+1}<1$ and \eqref{maeh^3} we obtain 
\begin{equation}
z_0 < \frac{\prod_{i=0}^{k}a_i}{1 + \sum_{i=0}^{k}\prod_{j=i}^{k}a_j} .
\end{equation}
``If'': As mentioned before, \eqref{boy2.1} and \eqref{boy4} imply
that the denominator in \eqref{boy3} is positive and greater $z_0$. Hence,
$0 < z_k < 1$, $k\in\N_0$. 
Furthermore, \eqref{boy6} shows that \eqref{boy3} implies \eqref{boy5}.
Now,
\begin{eqnarray}
\psi_k(z_{k+1}A_{k+1}) & \stackrel{\eqref{alder}}{=} & z_{k+1}[1-z_k]\psi_k(S_{k+1}/S_{k})A_{k} \\
\nonumber & \stackrel{\eqref{boy1}\&\eqref{boy5}}{=} & z_{k}A_{k}.
\end{eqnarray}
That means the strictly regular consumption process defined by \eqref{boy2} has the PHPP $\psi$.
  \end{proof}

For infinite time horizons, the number of 
possible regular consumption processes $X(S,\psi)$ for a given 
value process $S$ and a positively homogeneous projection property $\psi$
may be infinite -- in contrast to the uniqueness proved for the finite time case
(Theo.~\ref{theo_1}). 
The reason for this is the missing ``boundary condition'' \eqref{prop_2}.
For instance, any existing regular consumption solution $X$ in the infinite case 
can simply be rescaled
by a factor between zero and one and still be a regular solution for the same value process 
and the same property $\psi$, the only implication being that the money account $A$ would have
greater values than in the original case. 
In the case of Prop.~\ref{theo_3}, one could ask for the upper boundary of all possible 
$z_0$. By \eqref{boy4}, the answer is
\begin{equation}
\label{z_max}
z_0^{\text{max}}\; = \; \inf_{k\in\N_0} 
\left\{ \frac{\prod_{i=0}^{k-1}a_i}{1 + \sum_{i=0}^{k-1}\prod_{j=i}^{k-1}a_j} \right\} .
\end{equation}
All strictly regular consumption solutions $X(S,\psi)$ for
the given process $S$ and PHPP $\psi$ as stated in 
Prop.~\ref{theo_3} can therefore be obtained by
starting with any $z_0\in (0,z_0^{\text{max}})$, using \eqref{boy3} to obtain $z_k$ ($k>0$).
Whether $z_0^{\text{max}}$ itself can be used as a starting point depends on 
whether the infimum is actually reached in \eqref{z_max} (no), or not (yes).

\begin{example}
Let us consider the case of Prop.~\ref{theo_3} when $a_k=a$ for all 
$k\in\N_0$. In this case, \eqref{z_max} boils down to 
\begin{equation}
z_0^{\text{max}}\; = \; \inf_{k\in\N_0} 
\left\{ \frac{a^k}{\sum_{i=0}^{k}a^i} \right\} = \lim_{k\rightarrow\infty} 
\left( \sum_{i=0}^{k}a^{-i} \right)^{-1} = 1 - \frac{1}{a} .
\end{equation}
This makes sense for $a > 1$ only, since regularity is wanted.
From \eqref{boy5} we get that
\begin{equation}
z_k = 1 - \frac{1}{a} , \quad k\in\N_0.
\end{equation}
Therefore, the ``maximal'' solution ($z_0$ maximal) is a constant solution. 
Observe that $\psi_k(z_{k+1}A_{k+1}) = z_kA_k$ for identical $z_k$ ($k\in\N_0$)
implies
\begin{equation}
\psi_k(A_{k+1}) = A_k, \quad k\in\N_0.
\end{equation}
In other words, the money account at integer times, i.e.~the process $(A_k)_{k\in\N_0}$, 
has the positively homogeneous projection property $\psi$, too. 
In particular, if $\psi_k(\cdot) = \E[\cdot|\F_k]$ and $X$ is a martingale,
$(A_k)_{k\in\N_0}$ is a martingale, too. 
\end{example}


\section{Income drawdown}
\label{actuarial}

Income drawdown, sometimes also called {\em annuity drawdown} or {\em pension fund
drawdown}, was introduced in the UK in 1995 to provide annuity holders with better
income flexibility. 
The basic idea is the following. Instead of paying into a pension 
fund until retirement (e.g.~at age 65) and then immediately buying an annuity, the 
policyholder has the opportunity to start taking regular benefits from his/her 
accumulated funds already from the age of 50, or older (an increase to 55 is planned).
However, at the latest by the age of 75, an annuity has to be purchased from the remaining
funds to provide a secure lifetime income. In the UK, the monthly, quarterly or yearly
benefit during the drawdown period must be chosen between limits that are set by a 
government actuary each year. The policyholder can only vary the income within these limits.
More information on income drawdown can be found through the consumer information web pages
of the UK Financial Services Authority (FSA), or the web pages of the Government Actuary's 
Department (GAD). Also various private companies in the UK provide online information.

Let us assume that at time $k=0$ a life of age $x$ (e.g.~$x=55$) has the amount $S_0$ invested 
in a pension fund. Without consumption (income drawdown) the fund would develop like $S$,
where $S_0$ and $S$ are defined as before.
Assume that the policyholder wants to draw income yearly and
finally invests the remaining funds $A_K$ into an annuity at age $x+K$ (e.g.~$K=10$, $x+K=65$).
If we further assume that the policyholder draws income using a martingale
consumption strategy as in Example \ref{ex1}
($c=1$, ``consume every year what you can expect to consume the following year"),
then
\begin{equation}
\label{Ankara}
X_k = d\E[A_K|\F_k] , \quad k = 0, \ldots, K.
\end{equation}
The amount $A_K$ ($A_{10}$) left at the
end of the drawdown period is important since it is
used to purchase the obligatory annuity and therefore determines the amount of the annuity paid
later on. In international actuarial notation, $\ddot a_{x+K}$ ($\ddot a_{65}$) denotes the expected 
present value (and therefore actuarial price) of an 
annuity of one currency unit annually in advance for a life aged $x+K$ (65). Estimated values of
$\ddot a_{x+K}$ can be found in professionally used life tables, e.g.~$\ddot a_{65}= 13.666$
in the PMA92C20-table of the Actuarial Profession in the UK for a fixed effective annual interest 
rate of 4\%.
The annual amount of the annuity purchased at age $x+K$ for a sum
of $A_K$ is given by $B=A_K/\ddot a_{x+K}$. By \eqref{Ankara},
\begin{equation}
\label{stockbridge}
X_k = d \ddot a_{x+K}\E[B|\F_k] , \quad k = 0, \ldots, K.
\end{equation}
Once we have picked a value $d\in [0,1]$, the martingale strategy is fully determined.
For instance, under the assumptions of Prop.~\ref{relative2} one could then derive $X_0$, and
pursue the strategy. 
But what might be a reasonable choice of $d$? 
For example, $d=1/\ddot a_{x+K}$ would mean that a rate $X_k$ at any time $k$ is exactly the expected 
annual benefit of the annuity that will finally be bought at time $K$ (from the funds $A_K$). 

Especially the mentioned GAD, which has to
determine the upper limit of what a policyholder is allowed to consume each year, could make use of 
\eqref{stockbridge}. 
Obviously, $X_0$ depends on $d$. Provided the GAD decided on an upper 
consumption limit $L$, it is, if $L$ is not too large, possible to determine $d_L$ such that 
$X_0(d_L)\approx L$. 
By \eqref{stockbridge},
$L/(d_L \ddot a_{x+K})$ would then be the expected amount of the annuity, 
$\E[B]$, provided a ``smooth" consumption strategy
(martingale) was pursued from then on. The GAD could use $L/(d_L \ddot a_{x+K})$ as a benchmark value
and lower $L$ if $L/(d_L \ddot a_{x+K})$ was too small. The same procedure could be applied in the 
submartingale case $(c<1)$, allowing for consumption rates that grow on average.



\section{Smooth bonus for a closed with-profits fund}
\label{bonus}

An endowment policy with term $K$ years is a life insurance product that pays the policyholder 
the sum assured $(SA)$, e.g.~$SA=100,000$, at the end of the year of death if the policyholder dies 
before $K$, or it pays $SA$ on survival to $K$. We assume now that at time 0 $N_0$ of 
such policies are gathered in a closed with-profits fund. The fund is closed in the sense that 
no new policies are issued and no premiums are paid after time 0.
We denote the actuarial value (e.g.~the expected present value) of such a contract at time 0 for a 
life aged $x$ by $SA\times\Axn{x}{K}$ ($0<\Axn{x}{K}<1$). 
With this amount, we assume, the policy can be hedged sufficiently by buying corresponding securities, or even by buying reinsurance, i.e.~selling the liabilities
to another insurance company. Furthermore, assume that the funds belonging to one such policy are
$SA\times\Axn{x}{K} + B$ ($B>0$) at time 0. Since $B$ is not directly needed to secure the liabilities,
this amount is invested in a stochastic asset $R$, represented by a strictly positive stochastic process, 
$(R_t)_{0\leq t \leq K}$ $(R_0\equiv 1)$. 
The amount $B$ belongs to the insured and must therefore be redistributed over time.
One way to achieve this is to annually declare a bonus rate $b_k$ such that the current sum assured for 
the survivors at this time is increased by $b_k\times SA$.
Our goal is now to achieve that the process of bonus rates $b=(b_k)_{k = 0, \ldots, K}$ has a certain 
property, e.g.~is ``smooth" in the sense of being a martingale.
For this purpose we will apply Lemma \ref{rescaling}.

Let $N_k$ denote the number of survivors to time $k$. We make the technical assumption $N_k>0$.
Define now 
\begin{equation}
F_k = SA\times N_k \times \Axn{x+k}{K-k}, \quad k = 0, \ldots, K,
\end{equation} 
and 
\begin{equation}
S_k = \frac{N_0\times B \times R_k}{F_k},
\end{equation}
where we assume that $\Axn{x+k}{K-k}$ is strictly positive $(k = 0, \ldots, K)$.
Note that $\tilde{S}_k := F_kS_k = N_0\times B \times R_k$ is the value of the investment of 
the $N_0$ individual amounts $B$ in $R$ without consumption. $(\tilde{S}_t)_{0\leq t \leq K}$ is 
therefore from what the insurer can consume to pay bonus. Choose now a PHPP for $b$ consuming from 
$S$ (note: $S$ is artificial).
Assume now that $b_k$ is the bonus rate guaranteed at $k$. Then,
\begin{equation}
N_k\times (b_k\times SA)\times \Axn{x+k}{K-k} = F_kb_k = \tilde{b}_k
\end{equation} 
must be taken from the $R$-investment at time $k$ (identical to $\tilde{S}$ with consumption, 
or $\tilde{A}$)
to hedge or reinsure the additional insurance benefit $b_kSA$, that is guaranteed to all $N_k$
surviving policyholders by declaring
the bonus rate $b_k$. The problem is now to clarify whether this is possible at any $k$ without
ever getting into negative numbers. The answer is given by Lemma \ref{rescaling}.
Any $b$ regular w.r.t.~$S$ guarantees regularity of $\tilde{b}$ w.r.t.~$\tilde{S}$.
Therefore, $b$ defines a bonus strategy that is $(a)$ smooth, e.g.~when using a martingale with total
consumption at $K$,
and $(b)$ fair for the insured and the insurer, since every policyholder gets in expectation the same
amount paid as bonus, all money available for bonus (=$R$-investment)
is redistributed until $K$, and regularity ensures
that the insurance company can sufficiently hedge the liabilities evolving from this bonus
strategy.


\section{Conclusion}
\label{conclude}

This paper has introduced a general method to describe properties
of consumption processes for stochastic money accounts by means of certain positively homogeneous
operators (``positively homogeneous projection properties"). The main result, Theorem \ref{theo_1}, 
not only provided an existence theorem, 
but at the same time a backward-forward recursion method to explicitly calculate such strategies
in many cases. Relative consumption rates played a predominant role, as opposed
to absolute consumptions rates that were actually aimed for.
The properties seemed to be rather restrictive at first sight. However,
a result for finite spaces showed that almost any reasonable consumption strategy (i.e.~also 
such possibly obtained from an optimality criterion) 
can be described and constructed by these properties,
proving, in fact, how general the approach is. Furthermore, the conditional expectation and 
conditional quantile strategies that we described in examples are suitable for applications 
and are easy to communicate, even to non-mathematicians. 
The second part of the paper focused on the case when relative rates of consumption are
deterministic. In this case, we could derive much more explicit results that are particularly
suitable for applications with i.i.d.~growth rates for the original investment dynamics
(e.g.~log-L\'evy processes). Two numeric examples, binomial tree and geometric Brownian motion,
illustrated the results and the potential of the described methods. 
In particular, the potential to plan and provide ``smooth" or ``smoothly growing"
income under adverse market conditions should be emphasized.
We further showed that similar results and methods can be derived for
perpetual consumption problems.
Finally, the sections on income drawdown and on bonus for closed with-profits funds
showed applicability to actuarial problems.



%% file: picture1.tex
\begin{picture}(230,120)(0,0)

\put(0,0){\vector(4,0){235}}
\put(0,0){\line(0,4){85.9956}}
\put(0,103.3956){\vector(0,4){46.6044}}
\put(0,0){\dottedline{1}(0,85.9956)(0,103.3956)}
\multiput(0,-2.5)(60,0){4}{\line(0,6){5}}
\multiput(-2.5,0)(0,60){3}{\line(6,0){5}}

\put(0,0){\dottedline{1}(0,103.3956)(5,109.41912)(10,119.75784)(15,123.8424)(20,122.58264)(25,120.93792)(30,129.53496)(35,120.9888)(40,115.6392)(45,105.18144)(50,88.80336)(55,89.11056)(60,66.45672)(65,75.6684)(70,79.68768)(75,69.42312)(80,65.94792)(85,61.1292)(90,58.17288)(95,70.60896)(100,71.58432)(105,77.29392)(110,83.70864)(115,83.62992)(120,78.16272)(125,87.74376)(130,89.9028)(135,95.16384)(140,97.4064)(145,96.43584)(150,92.5608)(155,95.64504)(160,94.11384)(165,97.26552)(170,93.49464)(175,90.84504)(180,93.4296)(185,95.046)(190,99.024)(195,102.14592)(200,102.1164)(205,104.41176)(210,104.37048)(215,100.43616)(220,107.05512)(225,110.07072)}


\put(0,103.3956){\circle*{2}}
\put(-23,92){$X_0 \{$}

{\thicklines
\put(0,0){\drawline(0,85.9956)(5,91.0054477741026)(10,99.6043091340831)(15,103.00149613175)(20,101.953735713938)(25,100.585798555761)(30,107.736079738171)(35,100.628116179799)(40,96.1787773127677)(45,87.4809086814526)(50,73.8590251927161)(55,74.1145278284182)(60,55.2730049482957)}
}

\put(60,55.2730049482957){\circle*{2}}
\put(0,0){\dottedline{1}(60,55.2730049482957)(60,37.8730049482957)}
\put(37,43){$X_1 \{$}

{\thicklines
\put(0,0){\drawline(60,37.8730049482957)(65,43.1226471548643)(70,45.4131937140172)(75,39.5635259652617)(80,37.5830450327643)(85,34.83690579501)(90,33.1521292669366)(95,40.2393240514129)(100,40.7951717385447)(105,44.0490143755691)(110,47.7046976879855)(115,47.6598359652052)(120,44.5441346086934)}
}

\put(120,44.5441346086934){\circle*{2}}
\put(0,0){\dottedline{1}(120,44.5441346086934)(120,27.1441346086934)}
\put(97,31){$X_2 \{$}

{\thicklines
\put(0,0){\drawline(120,27.1441346086934)(125,30.4714118509808)(130,31.2211973290904)(135,33.0482368428346)(140,33.8270269170294)(145,33.4899735073501)(150,32.1442602648469)(155,33.2153466564862)(160,32.6835957282581)(165,33.7781024977708)(170,32.4685616538336)(175,31.5484158469938)(180,32.445974741365)}
}

\put(180,32.445974741365){\circle*{2}}
\put(0,0){\dottedline{1}(180,32.445974741365)(180,15.045974741365)}
\put(157,20){$X_3 \{$}

{\thicklines
\put(0,0){\drawline(180,15.045974741365)(185,15.3062810422797)(190,15.9469012260454)(195,16.4496576272775)(200,16.4449037037419)(205,16.8145502459763)(210,16.8079024829833)(215,16.1743165600591)(220,17.2402389762324)(225,17.7258735227793)}
}

\put(-2.5,-15){0}
\put(57.5,-15){1}
\put(117.5,-15){2}
\put(177.5,-15){3}
\put(235,-15){$t$}
\put(-10,-2.5){0}
\put(-30,57.5){2500}
\put(-30,117.5){5000}
{\thicklines
\put(0,0){\drawline(210,125)(230,125)}
}
\put(240,125){$A$}
\put(0,0){\dottedline{1}(210,140)(230,140)}
\put(240,140){$S$}

\end{picture}

%% file: picture2.tex
\begin{picture}(260,240)(0,0)



\put(0,175){
\put(0,-5){\vector(0,4){65}}
\multiput(-2.5,0)(0,20){3}{\line(6,0){5}}
\multiput(0,0)(0,20){3}{\dottedline{2}(0,0)(250,0)}

\drawline(0.0	,	20.0000	)(2.5	,	19.9639	)(5.0	,	19.5486	)(7.5	,	26.7360	)(10.0	,	25.5742	)(12.5	,	28.1475	)(15.0	,	30.2889	)(17.5	,	32.7492	)(20.0	,	24.2628	)(22.5	,	25.6616	)(25.0	,	25.7362	)(27.5	,	18.8801	)(30.0	,	22.1166	)(32.5	,	26.1581	)(35.0	,	27.1400	)(37.5	,	26.2304	)(40.0	,	25.0303	)(42.5	,	28.9106	)(45.0	,	29.1340	)(47.5	,	36.4574	)(50.0	,	39.2257	)(52.5	,	35.5559	)(55.0	,	34.0303	)(57.5	,	36.3983	)(60.0	,	35.4069	)(62.5	,	35.7752	)(65.0	,	33.2117	)(67.5	,	39.3907	)
(70.0	,	45.2372	)(72.5	,	52.1348	)(75.0	,	48.6945	)(77.5	,	49.1283	)(80.0	,	42.6392	)(82.5	,	42.3063	)(85.0	,	38.5725	)(87.5	,	37.3624	)(90.0	,	35.2205	)(92.5	,	29.2603	)(95.0	,	29.6055	)(97.5	,	24.8658	)(100.0	,	26.4302	)(102.5	,	21.0340	)(105.0	,	18.6938	)(107.5	,	25.0683	)(110.0	,	26.8642	)(112.5	,	29.1323	)(115.0	,	25.0241	)(117.5	,	29.1148	)(120.0	,	25.0781	)(122.5	,	25.6043	)(125.0	,	20.0062	)(127.5	,	16.5145	)(130.0	,	18.4913	)(132.5	,	15.5826	)(135.0	,	14.3876	)(137.5	,	12.9448	)(140.0	,	15.3434	)
(142.5	,	14.0128	)(145.0	,	17.6638	)(147.5	,	21.3488	)(150.0	,	23.0303	)(152.5	,	25.5133	)(155.0	,	29.4214	)(157.5	,	28.7953	)(160.0	,	22.6981	)(162.5	,	21.9916	)(165.0	,	18.3692	)(167.5	,	18.3598	)(170.0	,	18.2720	)(172.5	,	20.8176	)(175.0	,	15.9976	)(177.5	,	21.6637	)(180.0	,	20.8120	)(182.5	,	17.0138	)(185.0	,	16.9340	)(187.5	,	20.9777	)(190.0	,	15.3049	)(192.5	,	20.8090	)(195.0	,	24.4904	)(197.5	,	24.3803	)(200.0	,	21.3128	)(202.5	,	23.8481	)(205.0	,	26.0460	)(207.5	,	25.4659	)(210.0	,	25.1024	)(212.5	,	24.1132	)
(215.0	,	23.9629	)(217.5	,	23.0518	)(220.0	,	14.0046	)(222.5	,	14.7523	)(225.0	,	12.4847	)(227.5	,	12.0734	)(230.0	,	12.4532	)(232.5	,	10.7243	)(235.0	,	10.7342	)(237.5	,	10.3577	)(240.0	,	12.6936	)(242.5	,	13.5497	)(245.0	,	15.7091	)(247.5	,	14.5723	)(250.0	,	13.1028	)
\put(-35,-2.5){8,000}
\put(-40,17.5){10,000}
\put(-40,37.5){12,000}
\put(-15,60){$S_t$}
}

\put(0,95){
\put(0,-5){\vector(0,4){60}}
\put(-17.5,55){$X_k$}
\put(260,20){$\alpha=0.1$}
\multiput(-2.5,0)(0,20){3}{\line(6,0){5}}
\multiput(0,0)(0,20){3}{\dottedline{2}(0,0)(250,0)}
{\thicklines
\put(	0.0	,0){\line(0,4){	19.9829	}}
\put(	25.0	,0){\line(0,4){	23.5430	}}
\put(	50.0	,0){\line(0,4){	29.5792	}}
\put(	75.0	,0){\line(0,4){	35.5759	}}
\put(	100.0	,0){\line(0,4){	32.7823	}}
\put(	125.0	,0){\line(0,4){	34.3226	}}
\put(	150.0	,0){\line(0,4){	39.4001	}}
\put(	175.0	,0){\line(0,4){	40.9046	}}
\put(	200.0	,0){\line(0,4){	48.1011	}}
\put(	225.0	,0){\line(0,4){	48.9259	}}
\put(	250.0	,0){\line(0,4){	54.8796	}}
\put(	0.5	,0){\line(0,4){	19.9829	}}
\put(	25.5	,0){\line(0,4){	23.5430	}}
\put(	50.5	,0){\line(0,4){	29.5792	}}
\put(	75.5	,0){\line(0,4){	35.5759	}}
\put(	100.5	,0){\line(0,4){	32.7823	}}
\put(	125.5	,0){\line(0,4){	34.3226	}}
\put(	150.5	,0){\line(0,4){	39.4001	}}
\put(	175.5	,0){\line(0,4){	40.9046	}}
\put(	200.5	,0){\line(0,4){	48.1011	}}
\put(	225.5	,0){\line(0,4){	48.9259	}}
\put(	250.5	,0){\line(0,4){	54.8796	}}
}
\put(-18,-2.5){0}
\put(-30,17.5){500}
\put(-40,37.5){1,000}
}

\put(0,15){
\put(0,-5){\vector(0,4){60}}
\put(-17.5,55){$X_k$}
\put(260,20){$\alpha=1/3$}
\multiput(-2.5,0)(0,20){3}{\line(6,0){5}}
\multiput(0,0)(0,20){3}{\dottedline{2}(0,0)(250,0)}
{\thicklines
\put(	0.0	,0){\line(0,4){	32.3228	}}
\put(	25.0	,0){\line(0,4){	34.9731	}}
\put(	50.0	,0){\line(0,4){	40.3535	}}
\put(	75.0	,0){\line(0,4){	44.5732	}}
\put(	100.0	,0){\line(0,4){	37.7207	}}
\put(	125.0	,0){\line(0,4){	36.2696	}}
\put(	150.0	,0){\line(0,4){	38.2369	}}
\put(	175.0	,0){\line(0,4){	36.4569	}}
\put(	200.0	,0){\line(0,4){	39.3718	}}
\put(	225.0	,0){\line(0,4){	36.7783	}}
\put(	250.0	,0){\line(0,4){	37.8866	}}
\put(	0.5	,0){\line(0,4){	32.3228	}}
\put(	25.5	,0){\line(0,4){	34.9731	}}
\put(	50.5	,0){\line(0,4){	40.3535	}}
\put(	75.5	,0){\line(0,4){	44.5732	}}
\put(	100.5	,0){\line(0,4){	37.7207	}}
\put(	125.5	,0){\line(0,4){	36.2696	}}
\put(	150.5	,0){\line(0,4){	38.2369	}}
\put(	175.5	,0){\line(0,4){	36.4569	}}
\put(	200.5	,0){\line(0,4){	39.3718	}}
\put(	225.5	,0){\line(0,4){	36.7783	}}
\put(	250.5	,0){\line(0,4){	37.8866	}}
}
\put(-18,-2.5){0}
\put(-30,17.5){500}
\put(-40,37.5){1,000}
}

\put(0,-5){
\put(0,0){\vector(4,0){270}}
\multiput(0,-2.5)(25,0){11}{\line(0,6){5}}
\put(-2.5,-15){0}
\put(22.5,-15){1}
\put(47.5,-15){\dots}
\put(222.5,-15){9}
\put(247.5,-15){10}
\put(270,-15){$t (k)$}
}

\end{picture}